\documentstyle [prl,twocolumn,aps] {revtex}
\begin{document}
\draft
\title{ Generalized Fractional Statistics          }
\author{ G. Kaniadakis, A. Lavagno
              \thanks{e-mail: alavagno@polito.it}
             and P. Quarati}
\address{ Dipartimento di Fisica and INFM- Politecnico di Torino \\
Corso Duca degli Abruzzi 24, 10129 Torino, Italy \\ 
Istituto Nazionale di Fisica Nucleare, Sezioni di Cagliari e di Torino}
\maketitle
\begin {abstract} {\bf Abstract:} 
We link, by means of a semiclassical approach, 
the fractional statistics of particles obeying the Haldane 
exclusion principle to the Tsallis statistics and derive a 
generalized quantum entropy and its associated statistics.
\end {abstract}

The generalized non-extensive statistics and the fractional exclusion 
statistics have excited great interest because of the deep insights on 
the classical and quantum behavior of many different physical systems 
and because of the wide range of their applications.\\
The notion of generalized entropy, based on multifractal concepts, 
has been introduced by Tsallis \cite{ts1},  
 
\begin{equation}
S_q=\sum_{i=1}^W S_q^{\, i} \ \  \ \ \ \ \   
S_q^{\, i}=\frac{k_{_B}}{q-1} \, n_i (1-n_i^{q-1}) \ \ ,
\end{equation}
where $S_q^{\, i}$ is the entropy per state, 
$k_{_B}$ is the Boltzmann's constant, $W$ is the total number 
of possible microscopic configurations and $\{n_i\}$ are the 
associated probabilities.
The limit $q \rightarrow 1$ of $S_q$ yields the well-known 
Shannon entropy $S_1=-k_{_B} \sum_i n_i \log n_i$. For general 
values of $q$, $S_q$ satisfies the usual properties of non-negativity, 
equiprobability, irreversibility and concavity. 
One of the most important differences between $S_q$ and the 
Shannon entropy is that $S_q$, for $q \ne 1$, is non-extensive. 
Such a property plays a relevant r\^ole in many physical systems 
where long-range interactions are present.
Furthermore, it has been observed \cite{tsa10} that non-extensive 
behavior is strictly 
connected to quantum groups (q-deformation algebra, q-oscillators).
The generalized Tsallis statistics (TS) has recently received 
great attention; many authors have analyzed the thermodynamic 
properties of TS \cite{ram,pla3,tsa6} and  
the Tsallis equilibrium distribution has been used to study 
gravitational sistems \cite{pla}, 
solar neutrino problem \cite{ka4}
anomalous diffusion \cite{zan1,zan2}, optimization algorithms, 
statistical inference and probability theory \cite{pen}.
Stariolo \cite{sta} has shown that the TS can be derived as the 
asymptotic equilibrium distribution of a Langevin and a Fokker-Planck 
equation containing a properly defined generalized potential. 
Furthermore, 
considering a gas of non-interacting quantum particles in equilibrium 
condition, the 
Fermi-Dirac (FD) and the Bose-Einstein (BE) distributions have 
been generalized
{\it \`a la} Tsallis \cite{buyu}.

In his formulation of exclusion statistics, 
Haldane defined a generalized Pauli exclusion principle (HE) 
introducing the dimension $d_{_N}$ of the Hilbert space for 
single particle states as a finite and extensive quantity that 
depends on the number $N$ of particles contained in the 
system \cite{hal}. 
The exclusion principle implies that the number of available single 
particle states 
decreases as the occupational number increases 
\begin{equation}
\Delta d_{_N} = - g \, \Delta N  \ \ ,
\end{equation}
where $g$ is the parameter that characterizes 
the complete or partial action of 
the exclusion principle and   makes possible the interpolation between the BE 
($g=0$) and FD ($g=1$) statistics.
Following Haldane's formulation, several papers \cite{wu,wil,isa,raja,poly}
 have been devoted to study the equilibrium distribution function 
associated to this approach 
and the thermodynamic properties of many different physical systems. 
In particular, many authors have suggested the intrinsic connection 
between the fractional statistics and the interpretation of the 
fractional quantum Hall effect and anyonic physics within the 
Calogero-Sutherland model and the Luttinger model \cite{mur1,veigy,sen}.

Very recently, Rajagopal \cite{raja2} has considered the Tsallis's 
entropy by introducing the quantum degeneracy Haldane factor that 
defines the generalized exclusion principle and has deduced the 
resulting quantum distribution function when $q$ is very close to one.

In this paper we show that the extension of the TS to the quantum FD, 
BE and HE statistics can be obtained in a remarkably simple formalism 
by means of a kinetic approach, 
recently proposed by us \cite{ka2,lav}. This semiclassical formalism 
allows us to deduce the quantum distribution in a very natural mode 
without approximations.

It is well known that, in the decorrelation approximation, valid in the 
thermodynamic limit, the level mean occupation $n_i(t)=n(t,E_i)$ obeys 
the following master equation 
\begin{eqnarray}
\frac{ dn_i(t)}{dt}= &\pi & (t,E_{i+1}\rightarrow E_i)+
\pi (t,E_{i-1} \rightarrow E_i) \nonumber \\
-&\pi & (t,E_{i}\rightarrow E_{i+1})-\pi (t,E_{i} \rightarrow E_{i-1}) \ \ ,
\end{eqnarray}

\noindent
where $\pi(t,E_i \rightarrow E_{i+1})$ is the transition probability 
from the state $E_i$ to the state $E_{i+1}$. 
The Eq.(3) describes, 
in the nearest neighbor interaction approximation, 
the change of the population of the level $E_i$ due to the transitions 
to and from the levels  $E_{i\pm 1}$ . The particle kinetics is completely 
defined by means of the transition probability which we postulate 
\cite{ka2} given by 
\begin{equation}
\pi(t,E_i \rightarrow E_{i+1})= r(t,E_i,\Delta E) \, \varphi(n_i) \, 
\psi(n_{i+1}) \ \ ,
\end{equation}
where $\Delta E=E_{i+1}-E_i$, 
$r(t,E_i,\Delta E)$ is the transition rate from the state $E_i$ to the 
state $E_{i+1}$, $\varphi(n_i)$ is a function depending on the occupational 
distribution at the initial state $E_i$ and $\psi(n_{i+1})$ depends on the 
arrival state $E_{i+1}$. The functions $\varphi(n_i)$ and $\psi(n_i)$ must 
satisfy the following conditions: $\varphi(0)=0$ because the transition 
probability is equal to zero if the initial state is empty, $\psi(0)=1$ 
because the transition probability is not modified if the arrival state 
is empty. \\
The definition of the functions $\varphi(n_i)$ and $\psi(n_i)$ is a 
crucial point of this formalism since these two functions can inhibit 
or enhance the transition probability 
from a site to another one. In this sense Eq.(4) defines a generalized 
exclusion-inclusion principle which governs the particle kinetics.
Let us say that the function $\varphi(n_i)$ is proportional to the 
probability of finding in the state $E_i$ the occupation number $n_i$ 
and $\psi(n_i)$ is proportional to the probability of introducing an 
extra particle into a state with occupational number $n_i$. 
Therefore, these functions can be interpreted as the semiclassical 
analogous of the quantum creation and annihilation operators matrix 
elements 
in second quantization; this analogy can be shaped in the following
 expressions
\begin{equation}
\varphi (n_i) \propto \mid < n_i -1 \mid \hat{a}_{n_i} \mid n_i >\mid ^2 \ \ ,
\end{equation}

\begin{equation}
\psi (n_i) \propto \mid < n_i +1 \mid \hat{a}_{n_i}^{\dag} 
\mid n_i >\mid ^2 \ \ .
\end{equation}
Defining the particle current $j_i(t)=j(t,E_i)$ as 
\begin{equation}
j_i (t)=\left [ \pi(t,E_i \rightarrow E_{i+1}) -\pi(t,E_{i+1} 
\rightarrow E_i)\right ] \, \Delta E \ \ ,
\end{equation}
the master equation (3) can be written as 
a continuity equation for the distribution function $n_i(t)$ \cite{ka2}
\begin{equation}
\frac{ dn_i(t)}{dt}+\frac{ j_i (t)-j_{i-1} (t) }{\Delta E}=0 \ \ .
\end{equation}

In stationary conditions, in the limit $t\rightarrow \infty$, 
the particle current vanishes and posing $\varphi_i=\varphi(n_i)$, 
$\psi_i=\psi(n_i)$, 
$r_i=r(t \!\rightarrow\! \infty, E_i, \mid\!\!\Delta E\!\!\mid)$
we obtain the following balance equation
\begin{equation}
r_i \, \varphi_i \, \psi_{i+1}=r_{i+1} \, \varphi_{i+1} \, \psi_i \ \ .
\end{equation}

In this case, if we consider Brownian particles, 
the transition rate can be written as \cite{ka2} 
\begin{equation}
r_i=c \, e^{\beta E_i} \ \ ,
\end{equation}
where $\beta=1/(k_{_B}T)$, $E_i=\frac{1}{2} m v^2_i$ is the single 
particle kinetic energy
and $c$ is a constant.\\
Then, if we take into account Eq.(10),  we obtain from Eq.(9) 
\begin{equation}
\frac{\psi_i}{\varphi_i}=e^{\epsilon_i} \ \ ,
\end{equation}
where $\epsilon_i=\beta (E_i-\mu)$ is the dimensionless single particle 
energy defined up to an additive constant $\mu$, the chemical potential, 
that can be determined fixing the particle number of the system. \\
The above equation has a very simple and general form that defines 
intrinsically the statistical distribution and links two different 
quantities. In the r.h.s. the single particle energy appears and 
defines the particle interaction in the mean field approximation: 
infact, the single particle energy $\epsilon_i$ can be determined from 
the drift and the diffusion coefficients of the system \cite{ka2}. 
In l.h.s., the $\varphi_i$ and $\psi_i$ functions contain the classical or 
quantum behavior of the system.\\
The Eq.(11) defines a family of statistics for several choices of the 
function $\varphi_i$ and $\psi_i$. \\

In the classical case, the transition probability does not depend on the 
occupational distribution of the arrival site, hence $\varphi_i=n_i$, 
$\psi_i=1$ and 
we obtain the classical Maxwell-Boltzmann (MB) distribution 
\begin{equation}
n_i=e^{-\epsilon_i} \ \ .
\end{equation}

For $\varphi_i=n_i$ and $\psi_i=1+\kappa n_i$, we obtain the quantum 
fractional 
distribution \cite{ka2}
\begin{equation}
n_i=\frac{1}{\exp\epsilon_i - \kappa} \ \ ,
\end{equation}
where $\kappa\in [-1,1]$ is a parameter connected with the exchange 
statistical parameter $\alpha$ appearing in the quantum phase 
$e^{i\pi\alpha}$ \cite{nara}. 
For $\kappa=-1$, $0$ and $1$ one obtains the FD, MB and BE distribution, 
respectively.\\
In this case the function $\psi_i$ depends on the particle distribution of 
the arrival site. If $\kappa >0$ (boson-like particle) the transition 
probability is enhanced, if $\kappa <0$ (fermion-like particle) the 
transition probability is inhibited. Hence, the parameter 
$\kappa$ may be interpreted as the degree of indistinguishability 
or the degree of classicality of the particles under consideration.

As we have previously outlined, the HE statistics contains a 
generalized exclusion Pauli principle that influences the transition 
probability from a site to another. The function $\psi_i$ reflects 
these quantum properties. 
According to Karabali and Nair paper \cite{kara}, where the algebra 
of creation and annihilation 
operator of particles obeying the Haldane exclusion statistics  
has been considered, it 
appears consistent to postulate, in consequence of the semiclassical 
interpretation of the function $\psi_i$ given by Eq.s (5) and (6), the 
following relation 
\begin{equation}
\psi_i=[1-g n_i]^g \, [1+(1-g) n_i]^{1-g} \ \ .
\end{equation}
Inserting the above expression in Eq.(11) with $\varphi_i=n_i$, we obtain 
the HE distribution for the fractional exclusion statistics \cite{hal,wu,wil}
\begin{equation}
[1-g n_i]^g \, [1+(1-g) n_i]^{1-g} =n_i \, e^{\epsilon_i}\ \ .
\end{equation}
The Haldane's parameter $g$ defines the generalized exclusion Pauli 
principle \cite{hal} and for the particular cases $g=0$, $1$ one 
recovers the BE and FD distributions, respectively.\\

We have seen that the MB, FD, BE and HE statistics can be obtained 
from a semiclassical kinetic equation in the case of Brownian particles. 
Now we want to link the above statistical distributions to the Tsallis 
distribution 
\cite{ts1}
\begin{equation}
n_i= \{q [ 1-(1-q) \epsilon_i] \}^{1/(1-q)}\ \ .
\end{equation}
We wish to observe that the Druyvenstein distribution \cite{dru} can 
be seen as the limit $q\approx 1$ (weak collective interaction) of the 
Tsallis distribution. Infact, 
expanding the r.h.s. of Eq.(16) at the first order of ($q-1$), we obtain:
$n_i \simeq e^{-\epsilon_i-s \epsilon_i^2}$, where we have defined 
$s=(1-q)/2$. 
This description may play a relevant r\^ole in many physical systems; 
in Ref.\cite{ka4} we have applied this distribution to the solar core in 
order to solve the solar neutrino problem.

The TS describes the statistics of classical particles and can be 
obtained from Eq.(11) using $\varphi_i=n_i$, $\psi_i=1$ (classical 
transition probability) and posing, 
in place of $\epsilon$, the generalized energy \cite{sta}
\begin{equation}
\hat\epsilon_i=\frac{1}{q-1} \log\{q [1-(1-q) \epsilon_i]\} \ \ ,
\end{equation}
then Eq.(16) can be written as
\begin{equation}
n_i=e^{-\hat\epsilon_i} \ \ .
\end{equation}
In the limit $q\rightarrow 1$ we recover the MB distribution. \\
The functional modification of the single particle energy $\hat\epsilon$ 
may be interpreted as an effect of many-body effective interactions or of 
a collective interaction.

If we want to extend the classical TS to the quantum case, it is necessary 
to introduce 
the generalized energy $\hat\epsilon_i$ of Eq.(17), in the Eq.(11)
\begin{equation}
\frac{\psi_i}{\varphi_i}=e^{\hat\epsilon_i}\ \ .
\end{equation}
The functions $\varphi_i$ and $\psi_i$ suitable to the definition of the 
transition probability are strongly related to the quantum nature of the 
particles under consideration.\\

The TS of Eq.(16) has been obtained extremizing  the generalized entropy 
$S_q$ defined in Eq.(1), enforcing the constraints of fixed energy 
$E=\sum_i E_i \, n_i^q$
and of particle number $N=\sum_i n_i^q$
\begin{equation}
\delta S_q-\beta \, \delta E +\beta\mu \, \delta N =0 \ \ .
\end{equation}
The above variational problem produces the following equation for $n_i$
\begin{equation}
\frac{\partial}{\partial n_i} (S_q^{\, i}-\beta E_i \, n_i^q+\beta\mu \, 
n_i^q)=0 \ \ .
\end{equation}
Comparing Eq.s (19) and (21), it is possible to obtain a generalized 
Tsallis's entropy 
containing an exclusion-inclusion principle, defined by means of the 
functions $\psi_i$ and $\varphi_i$  as 
\begin{equation}
S_q^{\, i}=\frac{k_{_B}}{q-1} \left [ \int n_i^{q-1} \, 
\left (\frac{\psi_i}{\varphi_i}\right )^{q-1} dn_i -n_i^q \right ] \ \ .
\end{equation}
In the classical case $\varphi_i=n_i$, $\psi_i=1$ and Eq.(22) reduces to 
the Tsallis's entropy of Eq.(1).\\

Let us describe the quantum generalization of the classical TS; 
we fix $\varphi_i=n_i$, the exclusion-inclusion principle 
must be contained into the expression of the function $\psi_i$. \\
If we choose $\psi_i=1+\kappa n_i$, we obtain the following statistics
\begin{equation}
n_i=\frac{1}{ \{q[ 1-(1-q) \epsilon_i]\}^{1/(q-1)} -\kappa} \ \ ,
\end{equation}
which contains the generalized  BE ($\kappa=1$) and the FD ($\kappa=-1$) 
quantum version of the TS derived in Ref.\cite{buyu}. \\
In this case the entropy can be obtained inserting the expression 
$\psi_i=1+\kappa n_i$ into Eq.(22)
\begin{equation}
S_{q,\kappa}^{\, i}=\frac{k_{_B}}{q-1} \left [ \frac{(1+\kappa n_i)^q}
{\kappa q} -
 n_i^q \right ] - \frac{1}{\kappa q (q-1)} \ \ .
\end{equation}
In the limit $q\rightarrow 1$ of the above equation, one obtains the 
well known entropy for particles obeying BE ($\kappa=1$) and FD 
($\kappa=-1$) statistics.

Analogously, we observe that, inserting the expression of $\psi_i$ of Eq.(14), 
which describes the exclusion principle introduced by Haldane, 
in the entropy of Eq.(22), it is possible to write the generalized 
{\em Haldane-Tsallis exclusion} (HTE) entropy  
$S_{q,g}=\sum_i S^{\, i}_{q,g}$  
(which depends on the two parameters $q$ and $g$) in terms of a 
hypergeometric function $F$ \cite{abra} as 
\begin{eqnarray}
S_{q,g}^{\, i}=\frac{ k_{_B} } {1-q} &\Bigg [& 
\frac{ (1 - g n_i)^\alpha          }
          { \alpha \, g^{1-\beta}     } 
F(\alpha,\beta;\alpha+1;(1-g)(1-g n_i))  \nonumber \\
&+& n_i^q - \frac{1}{\alpha \, g^{1-\beta} } 
F(\alpha,\beta;\alpha+1;1-g)   \Bigg ] \ \ ,
\end{eqnarray}
where $\alpha=1+g (q-1)$, $\beta=(g-1)(q-1)$.\\
We observe that, in the limit $q \rightarrow 1$ of the Eq.(25), one 
obtains the von Neumann entropy per state of the HE statistics \cite{raja}.\\
The insertion of the entropy per state $S^i_{q,g}$ into Eq.(21) 
yields the HTE distribution statistics
\begin{equation}
[1-g n_i]^g \, [1+(1-g) n_i]^{1-g}=n_i \, \{q [ 1-(1-q) 
\epsilon_i]\}^{1/(q-1)} \ \ ,
\end{equation}
which, in the limit $q\rightarrow 1$, produces the HE distribution 
given by Eq.(15).

The statistics defined by means of Eq.(26) describes a system of 
quantum particles with an  intermediate behavior 
between BE and FD statistics and, 
at the same time, characterized by 
a non-extensive entropy. 

In conclusion, we have shown that the classical and the quantum TS for 
Brownian particles can be obtained as stationary states of a linear 
or a non-linear (in the quantum case)
master equation if an exclusion-inclusion principle is introduced 
in the expression of the transition probability.
The classical or the quantum nature of the particles is 
taken into consideration 
by the functions $\varphi_i$ and $\psi_i$, this one enhances or 
inhibits the transition from a state to another.
The use of this semiclassical approach allows us to acquire a quantum 
generalization of the TS and to derive a generalized entropy together 
its associated 
statistics, linking the quantum behaviour of particles obeying the 
fractional exclusion principle introduced by Haldane to the non-extensive 
behaviour of the TS.
Situations of this kind can be present in different fields of physics from 
quantum cosmology to condensed matter \cite{chia}. 
As observed in Ref.\cite{tsa10,bacry} the non-extensive behavior of the 
system could be 
explain (part of) the discrepancy at the origin of the dark matter. 
Furthermore, the quantum entropy of Eq.(22) and its associated distribution 
statistics, may play a relevant r\^ole in those systems with long range 
interaction where the quantum properties are not negligible as:
neutron stars, black holes, astrophysical dense plasma, quark gluon plasma.
We quote in addition, as possible candidates,  
stellar polytropes, plasma heated by inverse bremsstrahlung, non thermal 
components of ions in fusion reactors and laboratory plasma, gases of 
particles in condensed matter correlated by many-body forces or 
long-range forces.
Of course, actually, studies on the thermodynamic properties of systems 
of particles obeying the HTE statistics have not yet been accomplished, 
nevertheless we think that this will be a field of advanced research 
in the near future.


\begin{references}
\bibitem{ts1}
C. Tsallis, J. Stat. Phys. 52 (1989) 479. 
\bibitem{tsa10}
C. Tsallis, Phys. Lett. A 195 (1994) 329.
\bibitem{ram}
J. Ramshaw, Phys. Lett. A 175 (1993) 169.
\bibitem{pla3}
A.R. Plastino and A. Plastino, Phys. Lett. A 177 (1993) 177.
\bibitem{tsa6}
C. Tsallis, Physica A 221 (1995) 277.
\bibitem{pla}
A.R. Plastino, A. Plastino, Phys. Lett. A 174 (1993) 384.
\bibitem{ka4}
G. Kaniadakis, A. Lavagno and P. Quarati,  Phys. Lett. B 369 (1996) 308.
\bibitem{zan1}
P.A. Alemany and D.H. Zanette, Phys. Rev. E 49 (1994) R956.
\bibitem{zan2}
D.H. Zanette and P.A. Alemany, Phys. Rev. Lett. 75 (1995) 366.
\bibitem{pen}
T. Penna, Phys. Rev. E 51 (1995) R1.
\bibitem{sta}
D.A. Stariolo, Phys. Lett. A 185 (1994) 262.
\bibitem{buyu}
F. Buyukkilic, D. Demirhan and A. Gulec, Phys. Lett. A 197 (1995) 209.
\bibitem{hal}
F.D.M. Haldane, Phys. Rev. Lett. 66 (1991) 1529.
\bibitem{wu}
Y.S. Wu, Phys. Rev. Lett. 73 (1994) 922.
\bibitem{wil}
C. Nayak and F. Wilckez, Phys. Rev. Lett. 73 (1994) 2740.
\bibitem{isa}
S.B. Isakov, Phys. Rev. Lett. 73 (1994) 2150.
\bibitem{raja}
A.K. Rajagopal, Phys. Rev. Lett. 74 (1995) 1048.
\bibitem{poly}
A. P. Polychronakos, Phys Lett. B 365 (1996) 202.
\bibitem{mur1}
M.V.N. Murthy and R. Shankar, Phys. Rev. Lett. 72 (1994) 3629.
\bibitem{veigy}
A.D. de Veigy and S. Ouvry, Phys. Rev. Lett. 72 (1994) 600; 
Mod. Phys. Lett. B9 (1995) 271.
\bibitem{sen}
D. Sen and R.K. Bhaduri, Phys. Rev. Lett. 74 (1995) 3912.
\bibitem{raja2}
A.K. Rajagopal, Physica B 212 (1995) 309.
\bibitem{ka2}
G. Kaniadakis and P. Quarati, Phys. Rev. E 49 (1994) 5103.
\bibitem{lav}
G. Kaniadakis, A. Lavagno and P. Quarati, Nucl. Phys. B (1996) in press.
\bibitem{nara}
R. Acharya and P. Narayana Swamy, J. Phys. A: Math. Gen. 27 (1994) 7247.
\bibitem{kara}
D. Karabali and V.P. Nair, Nucl. Phys. B 438 (1995) 551.
\bibitem{dru}
M. Druyvenstein, Physica (Eindhoven) 10 (1930) 6; 1 (1934) 1003.
\bibitem{abra}
M. Abramowitz and A. Stegun, {\it Handbook of Mathematical Functions}, 
Dover Pub., New York (1972), page 263.
\bibitem{chia}
{\it Common Trends in Condensed Matter and High Energy Physics}, 
Nucl. Phys. B (Proc. Suppl.) 33C (1993)
edited by L. Alvarez-Gaum\'e, A. Devoto, S. Fubini, C. Trugenberger 
(North-Holland, Amsterdam, 1993).
\bibitem{bacry}
H. Bacry, Phys. Lett. B 317 (1993) 523.

\end{references}
\end{document}